\documentclass[12pt]{iopart}

\usepackage{iopams}  
\usepackage{graphicx}
\usepackage{xcolor}
\usepackage{float}
\usepackage{subcaption}

\usepackage{multirow}

\usepackage[
  backend=biber,
  style=numeric,
  sorting=none,
  doi=false,
  isbn=false,
  url=false,
  eprint=false,
  date=year
]{biblatex}
\AtEveryBibitem{\clearfield{title}}
\renewbibmacro{in:}{}
\begin{document}


\title[Causality of ultrafast photoionization from argon $3s$]{Causality of ultrafast photoionization from argon $3s$ using an \textit{ab initio} relativistic approach}

\author{Rezvan Tahouri$^1$ and Jan Marcus Dahlstr\"{o}m$^{1,\ast}$}

\address{$^1$Department of Physics, Lund University, SE-221 00 Lund, Sweden}
\ead{$^\ast$marcus.dahlstrom@fysik.lu.se}
\vspace{10pt}
\begin{indented}
\item[] January 2026
\end{indented}


\begin{abstract}
\noindent
We study real-time photoionization flux at the $3s$ Amusia-Cooper minimum (ACM) in argon using \textit{ab initio} simulations with the relativistic time-dependent configuration-interaction singles (RTDCIS) method in length (LG) and velocity (VG) gauges. A simple analytical model is used to interpret the results, and to construct Wigner delays and Wigner distributions for both gauges and relativistic channels of the photoelectron ($\epsilon p_j$  with $j=1/2$ and $3/2$). The two gauges are found to produce qualitatively different ionization dynamics, with LG having positive and VG having negative Wigner delays. The advancement of several femtoseconds, found for Wigner delays in VG, raises some concern for causality when atoms are ionized by attosecond pulses that are shorter than the absolute value of the Wigner delay. Reassuringly, numerical simulations of wave packets with RTDCIS show that the electrons behave in a causal way in both gauges. Weighted delays that take into account the temporal window of excitation (or the bandwidth of the pulses) are constructed from the Wigner distribution to reach agreement between the numerical simulations and our simple wave packet model. Furthermore, a strong effect of spin-orbit coupling of the photoelectron ($j$) is reported for ultrafast photoionization dynamics, and Schrödinger kitten and cat states are identified in the Wigner distributions as a result of the ACM. Our work paves the way for a deeper understanding of ultrafast photoionization and the role of causality in systems with strong electron-electron correlation effects. 
\end{abstract}

\maketitle


\section{Introduction}

In many-electron atoms, mean-field and correlation effects play key roles in shaping bound and continuum electronic states, and thus the atomic response to external perturbations \cite{RevModPhys.40.441}. 
One dramatic example of how the reshaping of electron orbitals can change the cross section (CS) in photoionization \cite{flugge_theory_1982} is the Cooper minimum (CM) from the $3p$ orbital in argon atoms. It was first predicted by Cooper in 1962, using frozen core orbitals from the Hartree-Fock (HF) approximation \cite{Cooper1962}, and it sharply contrasts with the monotonously decreasing CS of hydrogen \cite{BetheSalpeter1957} and the positive matrix elements of the lighter noble gas atoms \cite{Cooper1962,kennedy_photoionization_1972}. There is no such minimum from the $3s$ orbital in argon within the independent particle model. However, in 1972, it was predicted by Amusia~\cite{Amusia1972} that a minimum in the partial CS from $3s$ should appear due to correlation-induced coupling from the $3p^{-1}$ pathway to the $3s^{-1}$ ionization channel. This correlation-induced Amusia–Cooper minimum (ACM) in the partial CS of the $3s$ channel \cite{Amusia1972} was later confirmed by synchrotron measurements~\cite{mobus1993PRA3S}. Much theoretical effort has been made to understand the nature of this correlation. It is known that there is strong configuration interaction (CI) between the $3s^{-1}$ and $3p^{-2}nl$ configurations in argon, e.g by Multiconfiguration Hartree-Fock (MCHF) \cite{froese_fischer_mchf_2007}, with the single hole $3s^{-1}$ being only $\sim 70\%$ of the $^2S(3s3p^5)$ state of the corresponding excited argon ion \cite{lagutin_photoionization_1999,carette_multiconfigurational_2013}. The role of such CI effects on photoionization time delays was first studied by Carette, who anecdotally referred to it as the ``argon nightmare'' because MCHF could successfully reproduce Fano resonances in the $3p^{-1}$ channel, induced by $3s^{-1}$, but it was not possible at the time to reproduce the overall ACM structure in the $3s^{-1}$ channel \cite{carette_multiconfigurational_2013}. For this reason, much future work was conducted instead in the spirit of Amusia with the Random Phase Approximation with Exchange (RPAE) \cite{Guenot_2012_PRA_Photoemission,Dahlstrom2012PRA,kheifets_time_2013,dahlstrom_study_2014,Vinbladh_PhysRevA_2019,kheifets_wigner_2023}, its relativistic variant (RRPA) \cite{saha_relativistic_2014,kheifets_dipole_2015,jordan_spin_orbit_2017,vinbladh_relativistic_2022,deshmukh_photoionization_2022,baral_unusual_2022,baral_dramatic_2023,baral_cooper_2024}, and Time-Dependent Local Density Approximation (TDLDA) \cite{dixit_time_2013, magrakvelidze_attosecond_2015, pi_attosecond_2018}. Early work on correlation effect in photoionization time delays \cite{moore_time_2011} have also been performed with the R-Matrix with Time-dependence (RMT) \cite{brown_rmt_2020}, which now allows for {\it ab initio} studies of the associated multiphoton effects and polarization in atoms and molecules \cite{benda_analysis_2022,benda_dipole-laser_2024,benda_angular_2025}. Recently, experimental and theoretical efforts on the photoionization of $3s$ in argon \cite{Alexandridi2021PRR,Hans_Jakob_2024,sizuo_prl_2025,PhysRevA111052805} have demonstrated that effects beyond RPAE are in fact required to describe photoionization on the attosecond time scale from the highly correlated $3s$ orbital in argon. \\ 

The concept of quantum time delay was first introduced by Eisenbud \cite{Eisenbud1948} and Wigner \cite{Wigner1955} in the context of single-channel scattering, and was generalized to multichannel cases by Smith \cite{Smith1960}, and reviewed by de Carvalho \cite{DECARVALHO200283}. This quantity, known as the Eisenbud-Wigner-Smith (EWS) delay, or often simply the Wigner delay, was later extended to photoionization to interpret the delay of an electron wave packet in attosecond experiments ~\cite{Schultze2010Science,nagele_time-resolved_2011, Kluender2011PRL, Dahlstrom2012JPB, Pazourek2015RMP}. Wigner delay in photoionization is defined as the energy derivative of the dipole matrix element phase, revealing complementary information not directly provided by the CS, and has become experimentally accessible indirectly through attosecond interferometric techniques, such as RABBITT~\cite{PaulScience2001RABBIT,toma_calculation_2002, Mairesse2003Science,Guenot2012PRA,Gruson2016Science,isinger2017science} and the attosecond-streak camera~\cite{Schultze2010Science,nagele_time-resolved_2011,Pazourek2015RMP}. 
At the ACM, the dipole phase moves rapidly in the complex plane, because the dipole matrix element passes very close to zero \cite{Amusia1972}, and highly affects the Wigner delay \cite{Guenot_2012_PRA_Photoemission}. Photoionization time delay measurements across $3s$ ACM have been performed, giving a negative delay, which corresponds to an {\it advancement} of the electron wave packet compared to the case without correlation ~\cite{Alexandridi2021PRR,sizuo_prl_2025}. On the contrary, many-body theories, such as RPAE, predicted a positive value corresponding to \textit{delayed}. photoelectrons compared to the reference ones ~\cite{kheifets_time_2013}. Formal RPAE calculations are gauge invariant, leading to consistent Wigner delays independent of the gauge. However, two commonly used approximations make the RPAE depend on the choice of gauge \cite{Vinbladh_PhysRevA_2019}: experimental binding energy corrections, which is a numerically cheap way to go beyond Koopman's theorem; and truncation of the active core orbitals, which makes the many-body calculations easier to converge. 
With such common modifications, the positive delay still remained \cite{Dahlstrom2012PRA,dahlstrom_study_2014}. This puzzle lasted for more than a decade until a joint experimental and theoretical study \cite{sizuo_prl_2025} resolved the discrepancy by reproducing the negative Wigner delay observed in a high-resolution attosecond interferometry measurement using RPAE calculations that include coupling to shake-up channels. Inclusion of shake-up channels has proven influential in time delay by Isinger \textit{et al.} where the relative delay between the $2p$ and $2s$ channels in neon was investigated and the observed behavior was attributed to additional correlation-induced shake-up channels~\cite{isinger2017science}. In addition to opening ionic channels, the shake-up mechanism in argon $3s$ coherently shifts the phase of the primary ionization channel, leading to advancement, rather than delay, of photoelectrons from the $3s$ channel. Meanwhile, the use of Kramers–Kronig relations to connect CS to the Wigner delay has opened a new perspective on the dipole in the complex plane \cite{Hans_Jakob_2024}. Large values of photoelectron advancement open up for detailed studies about {\it causality} in ultrafast excitations, {\it e.g.} a photoelectron should not be created before an attosecond pulse hits the atom.
 \\ 

 The physical description of many-body systems typically requires different approximations depending on the observables under study. In photoionization, electron-electron correlation effects are sensitive to methodological choices in theoretical simulations, such as truncation of basis functions and the level of many-body excitations. 
 The choice of interaction Hamiltonian between light and matter then introduces a gauge dependence; in exact theories with complete basis sets, length (LG) and velocity (VG) gauges are equivalent, but approximate methods like time-dependent configuration interaction singles (TDCIS)~\cite{Krebs2014, Greenman2010, PhysRevA.106.043104, PhysRevA.89.033415, pabst_spinorbit_2014} or its relativistic extension (RTDCIS)~\cite{ZapataPhysRevA2022} lead to gauge dependent results.
 Self-consistent approaches are gauge invariant and naturally the preferred option \cite{sato_gauge-invariant_2018, sato_time-dependent_2023}, but gauge-dependent theories may still offer a possible way to find numerically efficient compromises for specific functions \cite{bertolino_thomasreichekuhn_2022}. As an example, RTDCIS \cite{ZapataPhysRevA2022} in LG has been employed to study spin polarization and hole alignment in the vicinity of spectral features such as Fano resonances, CM, and giant resonances \cite{tahouri_relativistic_2024, carlstrom_spin_polarization_prl2025}. 
 The LG interaction involves the electric field and its dipole operator, which is rigorously applied to perturbative interactions and strong coupling regimes \cite{kobe_gauge_1978}. For the single electron hydrogen atom in an intense low-frequency fields, the interaction incorporates the momentum operator, and the results are found to be more stable in VG~\cite{Han2010PRA}. The freedom of gauge choice has its roots in classical electrodynamics, where different potentials represent the same physical fields~\cite{Sakurai_Napolitano_2020}. In quantum simulations with incomplete basis sets, observables such as photoelectron CS or Wigner delays often differ between gauges. In this work, we investigate this gauge dependence in the context of the argon $3s$ photoionization across the ACM, using the RTDCIS method for a real-time and real-space study with relativistic many-electron wave packets generated by ultrashort laser fields. We then match RTDCIS simulations in both LG and VG with an analytical model to obtain simple expressions for the dipole matrix elements that include correlation effects. We identify how gauge choice significantly impacts the theoretical predictions of time delays for outgoing photoelectron wave packets and discuss the role of causality using Wigner distributions.   

 The article is organized as follows. In Sec.~\ref {sec:theory}, we present our theoretical approach with an overview of the RTDCIS method in Sec.~\ref{sec:num} and of the analytical model in Sec.~\ref{sec:ana}. In Sec.~\ref{sec:res}, we demonstrate the numerical results with CS in Sec.~\ref{sec:cs} and Wigner delays in Sec.~\ref{sec:wignerdelay}; and time-dependent relativistic flux from attosecond excitation in Sec.~\ref{sec:tdflux-as} and femtosecond excitation in Sec.~\ref{sec:tdflux-fs}. Sec.~\ref{sec:disc} is the discussion, the role of causality is studied using the Wigner distribution in Secs.~\ref{sec:caus-Wig} and \ref{sec:dir-cau}; and our final interpretation of the observed delay of the RTDCIS flux is given in Sec.~\ref{sec:interp}. In Sec.~\ref{sec:concl} we draw our conclusions.
 

\section{Theoretical framework}
\label{sec:theory}

\subsection{Numerical method}
\label{sec:num}

The RTDCIS method is designed to model ultrafast electron dynamics, incorporating relativistic effects such as spin-orbit coupling. The starting point is the time-dependent Dirac equation (TDDE):
\begin{equation}
    \label{eq:TDDE}
    \mathrm{i} \frac{d}{dt} \vert \Psi (t) \rangle = \left[ \hat{H} + \hat{V}(t) \right] \vert \Psi(t) \rangle,
\end{equation}
where $ \hat{H} $ is the field-free Hamiltonian, and $ \hat{V}(t) $ describes the interaction between the electrons and the external coherent field. The many-electron wavefunction ansatz is approximated by a linear combination of the Dirac-Fock ground state and singly excited configurations:
\begin{equation}
    \label{eq:RTDCIS_ansatz}
    \vert \Psi(t) \rangle = c_0(t) \vert \Phi_0^{\mathrm{DF}} \rangle + \sum_{a,p} c_a^p(t) \vert \Phi_a^p \rangle,
\end{equation}
where the singly excited configurations, $\vert \Phi_a^p \rangle = \hat{a}^{\dag}_p \hat{a}_a \vert \Phi_0^{\mathrm{DF}}\rangle$, describe an electron excited from an occupied orbital $ a $ to an unoccupied (virtual) orbital $p$, and the ground state, $\vert \Phi_0^{\mathrm{DF}} \rangle = \hat{a}^{\dag}_N ... \hat{a}^{\dag}_c \hat{a}^{\dag}_b \hat{a}^{\dag}_a \vert 0 \rangle$, builds upon the vacuum state $\vert 0 \rangle$. Each orbital $\vert i \rangle$ in this basis is a four-component Dirac spinor, obtained from the solution of the one-electron mean field eigenvalue Dirac-Fock equation:
\begin{equation}
    \label{eq:DF_orbital}
    \hat{h}_0^{\mathrm{DF}} \vert i\rangle = \varepsilon_i \vert i \rangle,
\end{equation}
where $\hat{h}_0^{\mathrm{DF}}$ includes the relativistic kinetic energy, nuclear attraction, and an effective
one-electron operator called the Dirac-Fock potential, $\hat{\nu}^{\mathrm{DF}}$, and is written as
\begin{equation}
    \hat{h}_0^{\mathrm{DF}} = c\,\boldsymbol{\alpha} \cdot \mathbf{p} + \beta c^2 - \frac{Z}{r} + \hat{\nu}^{\mathrm{DF}}.
\end{equation}
Here, $c$ is the speed of light, $\boldsymbol{\alpha}=(\alpha_{x},\alpha_{y},\alpha_{z})$ and $\beta$ are Dirac matrices, $Z$ is the nuclear charge, and $\mathbf{p} = -\mathrm{i} \nabla$ and $r$ are the electron momentum operator and position, respectively. In spherical polar coordinates, the orbital $\vert i \rangle$ is given by

\begin{equation} \label{eq: spinor}
  \langle \mathbf{r} \vert i \rangle 
  \equiv \phi_{n,\kappa,m_j}(\mathbf{r})
  = \frac{1}{r}
  \left(
  \begin{array}{c}
    P_{n,\kappa}(r)\,\chi_{\kappa,m_j}(\Omega) \\[6pt]
    \mathrm{i}\,Q_{n,\kappa}(r)\,\chi_{-\kappa,m_j}(\Omega)
  \end{array}
  \right)
\end{equation}
where $P_{n,\kappa}(r)$ and $Q_{n, \kappa}(r)$ are the radial large and small components, respectively, $\chi_{\pm \kappa,m_j}(\Omega) $ are the spin-angular components, $\Omega$ represents $\theta$ and $\phi$ coordinates, and $\{ n, \kappa, m_j \}$ are quantum numbers \cite{grant_relativistic_2007, lindgren_rosen_relativistic_1975}. \\

The total field-free Hamiltonian for the system is written as:
\begin{equation}
    \label{eq:H_full}
    \hat{H} = \hat{H}_0^{{\mathrm{DF}}} + \hat{H}_1 - E_0^{\mathrm{DF}}.
\end{equation}
Here, $ \hat{H}_0^{\mathrm{DF}} $ is the sum of all one-electron Dirac-Fock operators acting independently on each electron:
\begin{equation}
    \hat{H}_0^{{\mathrm{DF}}} = \sum_{i=1}^{N} \hat{h}_0^{\mathrm{DF}}(i),
\end{equation}
where $N$ is the total number of electrons. Then, $ \hat{H}_1 $ is the difference between the Coulomb interaction and the Dirac-Fock potential: 
\begin{equation}
    \hat{H}_1 =  \sum_{i=1}^{N} \left( - \hat{\nu}^{\mathrm{DF}}(i) + \frac{1}{2}\sum_{j>i}^{N} \frac{1}{|\mathbf{r}_i - \mathbf{r}_j|} \right).
\end{equation}
The constant $ E_0^{\mathrm{DF}} $ is the energy of the Dirac-Fock ground state and is subtracted to become the zero energy reference. The interaction Hamiltonian in the length gauge is given by
\begin{equation}
    \label{eq:Vt}
    \hat{V}_{\mathrm{LG}}(t) = \sum_{i=1}^{N} E(t) \hat{z}(i),
\end{equation}
where $ E(t) $ is the electric field of the coherent field, and is assumed to be linearly polarized along the $ z $-axis, and $ \hat{z} $ is the $z$-component of the dipole operator. Alternatively, in the velocity gauge, the interaction is expressed in terms of the vector potential $ A(t) $, which is related to the electric field by $ E(t) = -d A(t)/d t $ and the $z$-component of the momentum operator, $\hat{\alpha}_z$ as 
\begin{equation}
    \hat{V}_{\mathrm{VG}}(t) = \sum_{i=1}^{N} c A(t) \hat{\alpha}_z(i),
\end{equation}
which mixes the large and small components. 
By inserting the wavefunction ansatz, Eq. (\ref{eq:RTDCIS_ansatz}), in TDDE, Eq. (\ref{eq:TDDE}), one obtains the RTDCIS equations of motion for $c_{0}(t)$ and $c_{a}^{p}(t)$ \cite{ZapataPhysRevA2022}. Such equations describe how the laser field affects the transitions between the ground and excited states, and how the excited states interact with each other because of the electron-electron interaction at the level of CIS. 


\noindent In quantum mechanics, probability flux is expressed as \cite{Sakurai_Napolitano_2020}:

\begin{equation}
\mathbf{j}(\mathbf{r},t) = -\frac{\mathrm{i}}{2}\Bigl[\Psi^*(\mathbf{r},t)\nabla\Psi(\mathbf{r},t) - \nabla\Psi^*(\mathbf{r},t)\Psi(\mathbf{r},t)\Bigr].
\end{equation}
Within the RTDCIS framework, time-dependent photoelectron fluxes are obtained using the relativistic many-electron solution of the TDDE, Eq. (\ref{eq:TDDE}), with time-dependent 4-component wave functions Eq.(\ref{eq: spinor}). In practice, we are interested in photoelectrons leaving a spherical volume of radius $R$, which leads to an integrated expression for the outgoing flux as

\begin{equation} \label{eq: rtdcis flux}
{\cal J}_{\kappa;h}(R,t) = -\mathrm{i}c \Bigl[ {P}^*_{\kappa;h}(R,t){ Q}_{\kappa;h}(R,t) - { Q}^*_{\kappa;h}(R,t){ P}_{\kappa;h}(R,t)\Bigr],
\end{equation}
which provides the photoelectron flux ionized from hole $h$ with different values of the relativistic quantum number $\kappa$ of the photoelectron. 

\subsection{Analytical model}
\label{sec:ana}

To interpret the results of RTDCIS near the $3s$ ACM of argon, we use a model for the dipole matrix element as a function of frequency \cite{sizuo_prl_2025}, 
\begin{eqnarray} 
& z_\pm(\omega)= z_0(\omega) \left[1-\varkappa \mathrm{e}^{\pm \mathrm{i} \Delta\varphi} \arctan\left(\frac{\omega-\epsilon_z}{\Delta \epsilon_z}\right) \right],\label{eq: dipole model} \\
& z_0(\omega)=a \omega^{-b}\mathrm{e}^{\mathrm{i} \sigma_1(\omega)}\theta(\omega-I_p), \label{eq: z_0} 
\end{eqnarray} 
which separates an uncorrelated background from a correlation-induced term given by an inverse tangent function to simulate the effect of a virtual transition near the $3p$ CM. We note that this model is \textit{ad hoc} and based on physical intuition rather than a formal derivation; however, the same model has recently proven useful for interpreting Wigner delay behavior in non-relativistic cases \cite{sizuo_prl_2025}. In this equation, the slowly varying part of the dipole element is  $z_0(\omega)$, $\varkappa$ is the strength of the correlation, and $\Delta \varphi$ is the relative phase of the correlation term. The negative sign for the correlated term comes from the combined effect of two $- \mathrm{i}$ phase factors: one from the scattering phase and another from higher-order perturbation theory, giving $(-\mathrm{i})^2 = -1$ in the virtual transition path. In our model, $\epsilon_z$ and $\Delta \epsilon_z$ are the frequency and width of the virtual $3p$ CM that is transferred to the $3s$ channel by interchannel coupling. The uncorrelated dipole matrix element, $z_0(\omega)$ depends on constants $a$ and $b$, and $\sigma_1(\omega)=\arg \left[ \left( \Gamma(2-\mathrm{i}/k(\omega)\right)\right]$ which is the Coulomb phase for a p-wave ($\ell=1$) with $\Gamma(x)$ being the Gamma function, $k(\omega)=\sqrt{2(\omega-I_p)}$ the momentum of photoelectron, and $I_p=1.2865\,\mathrm{au}=35.007\,\mathrm{eV}$, the ionization energy of the $3s$ orbital in RTDCIS. Then the CS in the length and velocity gauges for the relativistic case is written respectively as: 
\begin{equation} \label{eq:cs_lg}
\sigma_{\mathrm{LG}}(\omega)=4\pi^2 \alpha \omega | \hat{z}(\omega)|^2 \times a_0^2[\mathrm{Mb}],
\end{equation}
\begin{equation} \label{eq:cs_vg}
\sigma_{\mathrm{VG}}(\omega)=\frac{4\pi^2 \alpha  | c \hat{\alpha}_{z}(\omega)|^2}{\omega} \times a_0^2[\mathrm{Mb}],
\end{equation}
where $\alpha$ is the fine structure constant, and $a_0^2[\mathrm{Mb}]=28.0028$ is the conversion factor from atomic units to mega barns (Mb). The relativistic matrix element $c\hat{\alpha}_{z,\pm}(\omega)$ is modeled in the same way as $z_{\pm}(\omega)$ using Eq. (\ref{eq: dipole model}). Model parameters are determined separately in LG and VG by fitting the CS to the corresponding RTDCIS results. The CS does not depend on the sign of the correlation phase, and we can only determine $|\Delta \varphi|$ by this fitting \cite{sizuo_prl_2025}. The sign of the relative phase remains ambiguous when computed with CS in Eqs. (\ref{eq:cs_lg}) and (\ref{eq:cs_vg}). 

Once the parameters are fitted by the CSs, we evaluate the Wigner delay, which is a measure of the energy dependence of the ionization phase $\phi=\mathrm{arg}(z_\pm(\omega))$. Being defined as $\tau_{\mathrm{EWS}}=d\phi/dE$, the Wigner delay is highly sensitive to different signs $\pm|\Delta \varphi|$. \\
To study the real time dependence of the photoionization process, we build up coherent radial wave packets (for p-waves), using the dipole element in Eq. (\ref{eq: dipole model}) as
\begin{equation} \label{eq:wave packet}
\Psi_{\pm}(R, t) \propto \int d\omega \, z_{\pm}(\omega) g_{\omega_0,\tau}(\omega) \mathrm{e}^{\mathrm{i} [k(\omega)  R - (\omega - I_p) t +\ln (2k(\omega) R)/k(\omega)]}.
\end{equation}
The uncorrelated wave packet, $\Psi_0(R, t)$, is formed in the same way, but using $z_0(\omega)$ instead of $z_\pm(\omega)$. The spectral envelope $g_{\omega_0,\tau}(\omega)$ is taken to be a Gaussian function in VG, defined as:

\begin{equation}
g^\mathrm{VG}_{\omega_0,\tau}(\omega) = \exp \left[-\frac{\tau^2}{8 \ln 2} (\omega - \omega_0)^2 \right],
\end{equation}
where the central energy is $\omega_0$. In LG the filter is computed from the time derivative of the vector potential in VG, yielding $g^\mathrm{LG}_{\omega_0,\tau}(\omega) = \mathrm{i}\omega g^\mathrm{VG}_{\omega_0,\tau}(\omega)$.
 Time-dependent radial fluxes of outgoing photoelectrons at radius $r=R$ are obtained from the analytical model using 
 \begin{equation} \label{eq:flux}
{\cal J}_\pm(R,t) =\mathrm{i}\hbar/2m_e\big[\Psi_\pm^\ast\partial_r\Psi_\pm-\Psi_\pm\,\partial_r\Psi_\pm^\ast\big]_{r=R} 
\end{equation}
  Once the model parameters are obtained by fitting to RTDCIS results, we study the wave packets for both gauges in the time domain. This approach enables us to investigate the effect of gauge choices on the physics in the frequency and time domains.\\
To have a simultaneous visualization of the temporal and spectral characteristics of the photoionization process, the time–energy Wigner distribution function ${\cal W}(\omega,t)$ is employed. It is defined as \cite{cohen_time_frequency_1989}:
\begin{equation}
    {\cal W}(\omega,t) = \frac{1}{\pi} \, \mathrm{Re} \left[
        \int_{-\infty}^{\infty}
        \mathcal{Z}^*(\omega+\omega') \,
        \mathcal{Z}(\omega - \omega') \,
        e^{2 i t \omega'} \, d\omega'
    \right],
    \label{eq:wigner distribution}
\end{equation}
and has been used to study wave packets in photoionization previously \cite{sizuo_prl_2025,busto_timefrequency_2018}. 
Here, $\mathcal{Z}(\omega)$ is the \textit{effective} dipole amplitude computed as $\mathcal{Z}(\omega) = z(\omega) g_{\omega_0,\tau}(\omega)$ to filter the dipole amplitude with a Gaussian function corresponding to the same time duration as in the time-dependent flux calculations. 


\section{Results}
\label{sec:res}
\subsection{Cross section of model fitted to RTDCIS}
\label{sec:cs}
First, we perform correlated calculations using RTDCIS by including only $3s$ and $3p$ orbital couplings in VG and LG for different central frequencies with a pulse duration $12$ fs. Second, we optimize the model to reproduce the $3s$ partial CSs from those numerical simulations, obtaining the following values for the parameters in the Eqs (\ref{eq: dipole model}) and (\ref{eq: z_0}):

\begin{table}[H]
\centering
\caption{Model parameters obtained by fitting the analytical partial CS of $3s$ to the numerical ones from RTDCIS in LG and VG.}
\begin{tabular}{lcccccc}
\br
 & $\epsilon_z$ (eV) & $\Delta \epsilon_z$ (eV) & $\varkappa$ & $\Delta\varphi$ (rad) & $a$ & $b$ \\
\mr
LG  & 40.27 & 65.03 & 1.30 & 0.04 & 0.31 & 0.80 \\
VG   & 38.64 & 43.53 & 3.69 & 0.05 & 0.32 & 0.23 \\
\br
\end{tabular}
\label{tab:modelparams}
\end{table}
\noindent In LG, the virtual $3p$ CM is at a higher energy ($\epsilon_z$) with a broader spectrum ($\Delta \epsilon_z$) compared to VG. The correlation is stronger in VG because $\varkappa_{\mathrm{VG}} \approx 3 \times \varkappa_{\mathrm{LG}}$, while the correlation phases are similar, both with rather small magnitudes; around $0.8\%$ of $2\pi$ in VG and $0.6\%$ of $2\pi$ in LG. \\
\begin{figure}[htbp!]
    \centering
    \begin{subfigure}[t]{0.75\textwidth}
        \centering
        \includegraphics[width=\textwidth]{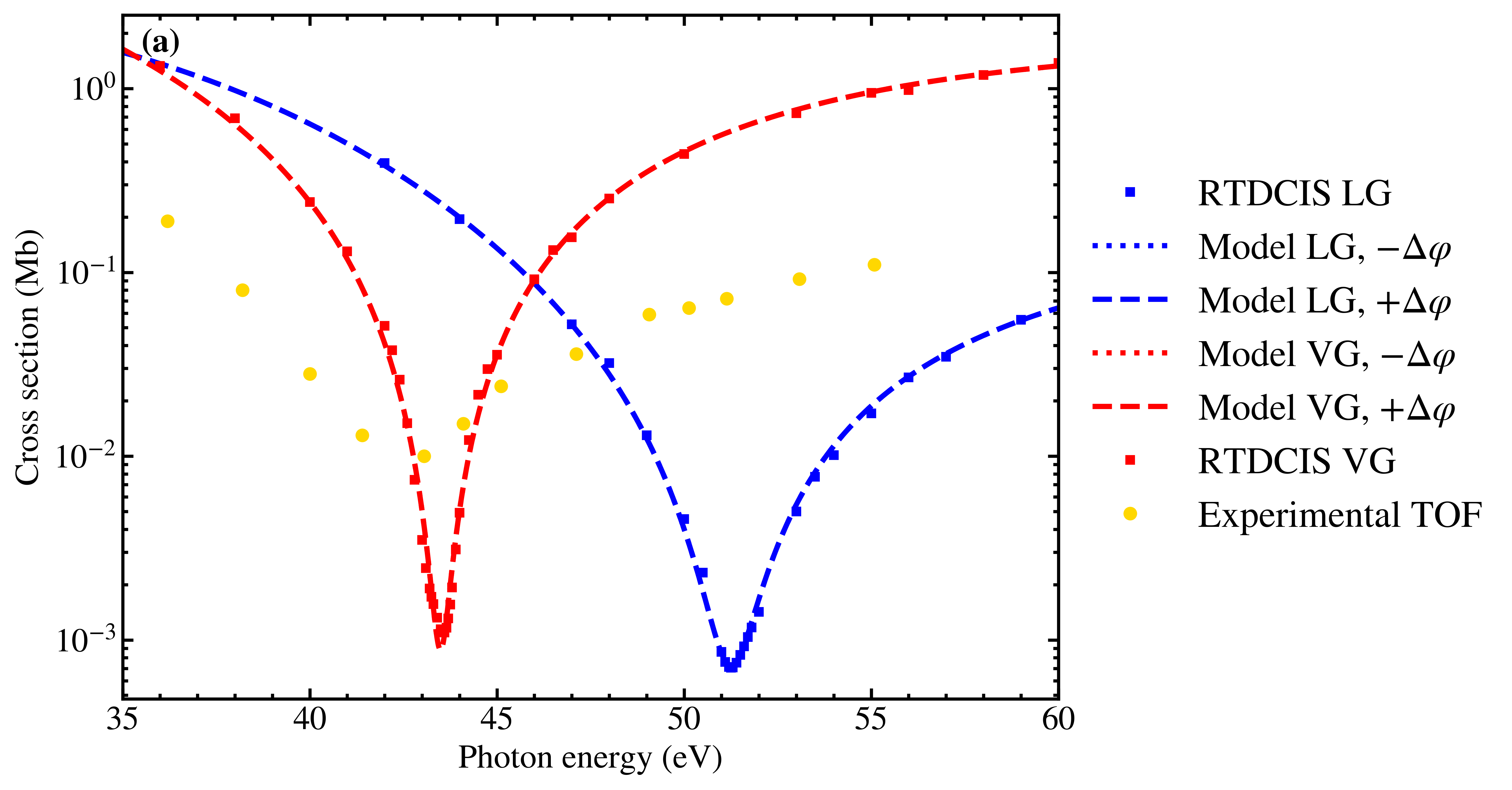}
        
    \end{subfigure}
    \\
    
    \begin{subfigure}[t]{0.49\textwidth}
        \centering
        \includegraphics[width=\textwidth]{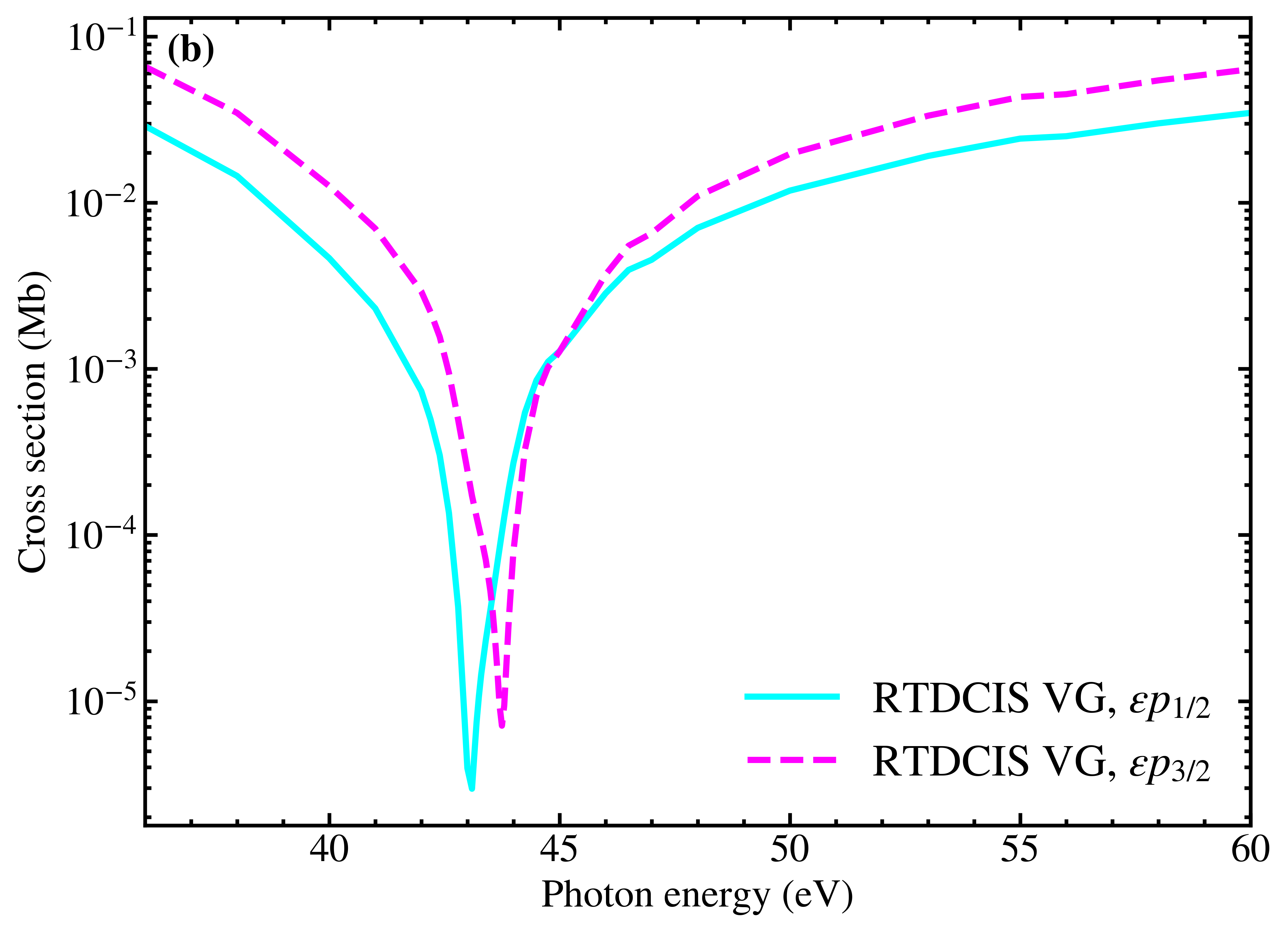}
        
    \end{subfigure}
    \hfill
    \begin{subfigure}[t]{0.49\textwidth}
        \centering
        \includegraphics[width=\textwidth]{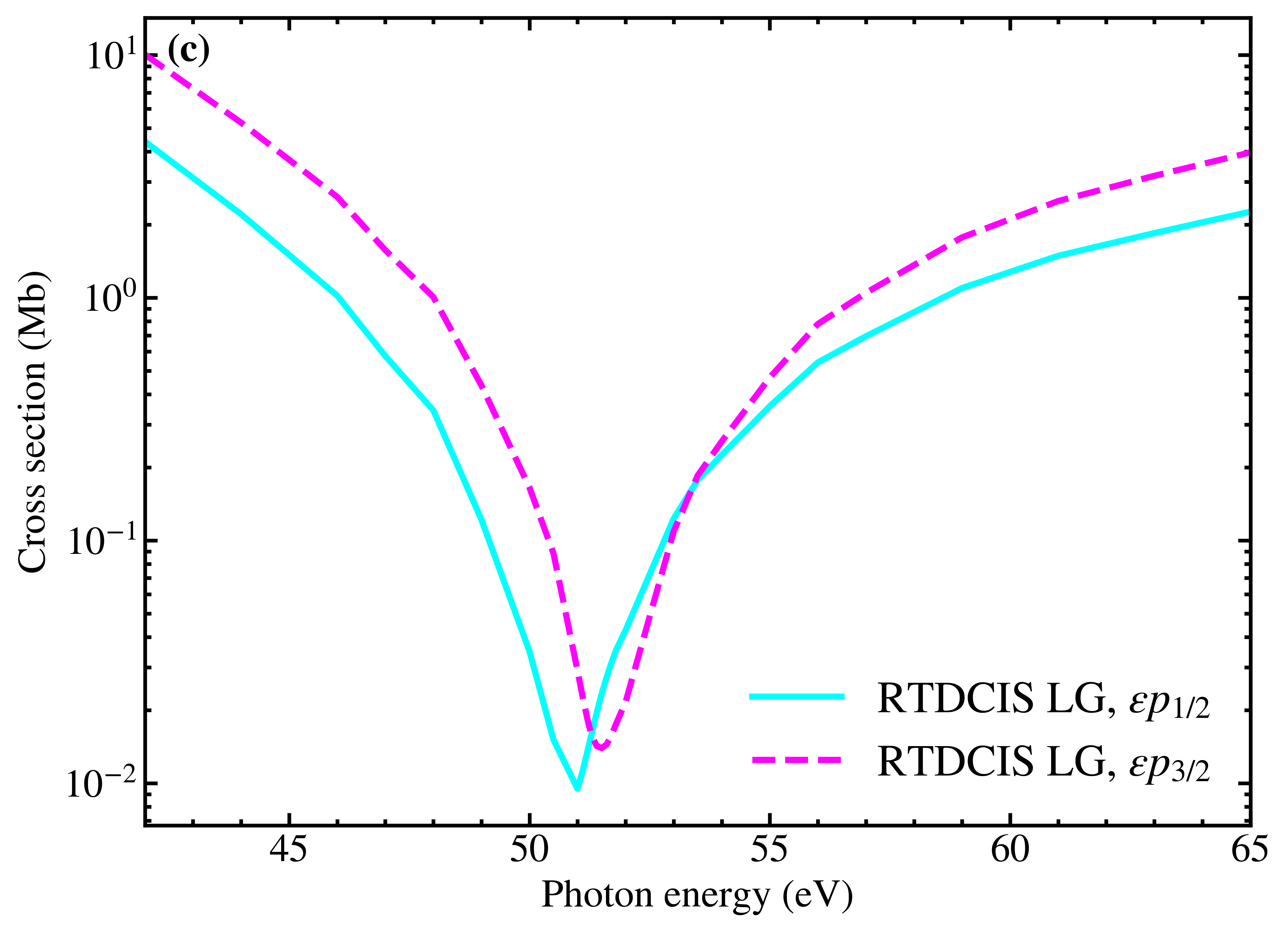}
        
    \end{subfigure}
    \caption{(a) Photoionization CS for the argon $3s$ orbital calculated using RTDCIS (squares) and analytical model (curves) in LG (blue) and VG (red), together with experimental TOF measurements (orange circles) from Ref.\cite{mobus1993PRA3S}. Panels (b) and (c) show the partial CS for the $3s_{1/2} \rightarrow \epsilon p_{1/2}$ (cyan curves) and $3s_{1/2} \rightarrow \epsilon p_{3/2}$ (magenta curves) relativistic channels in VG and LG, respectively.}
    \label{fig:cs_lg_vg}
\end{figure}

In Fig.\ref{fig:cs_lg_vg}(a), the $3s$ partial CS of argon is shown both experimentally from Ref. \cite{mobus1993PRA3S} using time-of-flight (TOF) measurements, and theoretically computed using RTDCIS in LG and VG, along with the corresponding fitted curves from the analytical model, Eqs. (\ref{eq:cs_lg}) and (\ref{eq:cs_vg}). As already mentioned, the sign of the correlation phase is ambiguous at the CS level, and that is why for each numerical result there are two corresponding fits of the model with opposite signs, but the same $|\Delta \varphi(\omega)|$. Comparing the two gauges, LG locates the ACM at $51.25$ eV and VG at $43.6$ eV. Both gauges have similar CS before the ACM (at $35$ eV), while after ACMs (at $60$ eV), the CS of VG is almost one order of magnitude larger than that of the LG. The experimental data have a broader structure compared to CS predicted by either gauge, and the minimum aligns more closely with that of VG. It should be mentioned that RPAE produces a better match with the CS, {\it c.f.} Ref.\cite{dahlstrom_study_2014}. This indicates that neither of the gauges in RTDCIS reproduces the experimental results with quantitative accuracy, but both reproduce qualitatively the ACM phenomenon. \\
\noindent Figures \ref{fig:cs_lg_vg} (b) and (c) depict partial CS of the two of the relativistic channels, namely $3s_{1/2}\rightarrow \epsilon p_{1/2}$ and $3s_{1/2} \rightarrow \epsilon p_{3/2}$, for VG and LG, respectively. In VG, the ACM for $\epsilon p_{1/2}$ is at $43.1$ eV and for $\epsilon p_{3/2}$ is at $43.75$ eV, so the total ACM at 43.6 eV falls noticeably closer to the $j = 3/2$ contribution. In LG, the minimum of $\epsilon p_{1/2}$ channel is at $51$ eV while the minimum of $\epsilon p_{3/2}$ is at $51.5$ eV, and the total minimum of $3s$ is between these values at $51.25$ eV. Additionally, the relativistic channel resolved CS in VG has a minimum of two orders of magnitude deeper than that of LG and the total CS in either gauge. This relativistic splitting of CM was recently shown by Baral {\it et al.} to be huge for high-$Z$ atoms using the RRPA \cite{baral_unusual_2022}, with angular distribution asymmetry parameters deviating from its non-relativistic value of 2 \cite{baral_dramatic_2023}. Here, we raise the question of whether the more subtle relativistic energy shifts of the ACM in $3s$ of argon, see panels (b) and (c) in Fig.~\ref{fig:cs_lg_vg}, can affect the behavior of the time-dependent fluxes when using a pulse tuned to the frequency of the total ACM in panel (a). \\
\begin{figure}[htbp!]
    \centering
    \includegraphics[width=0.6\textwidth]{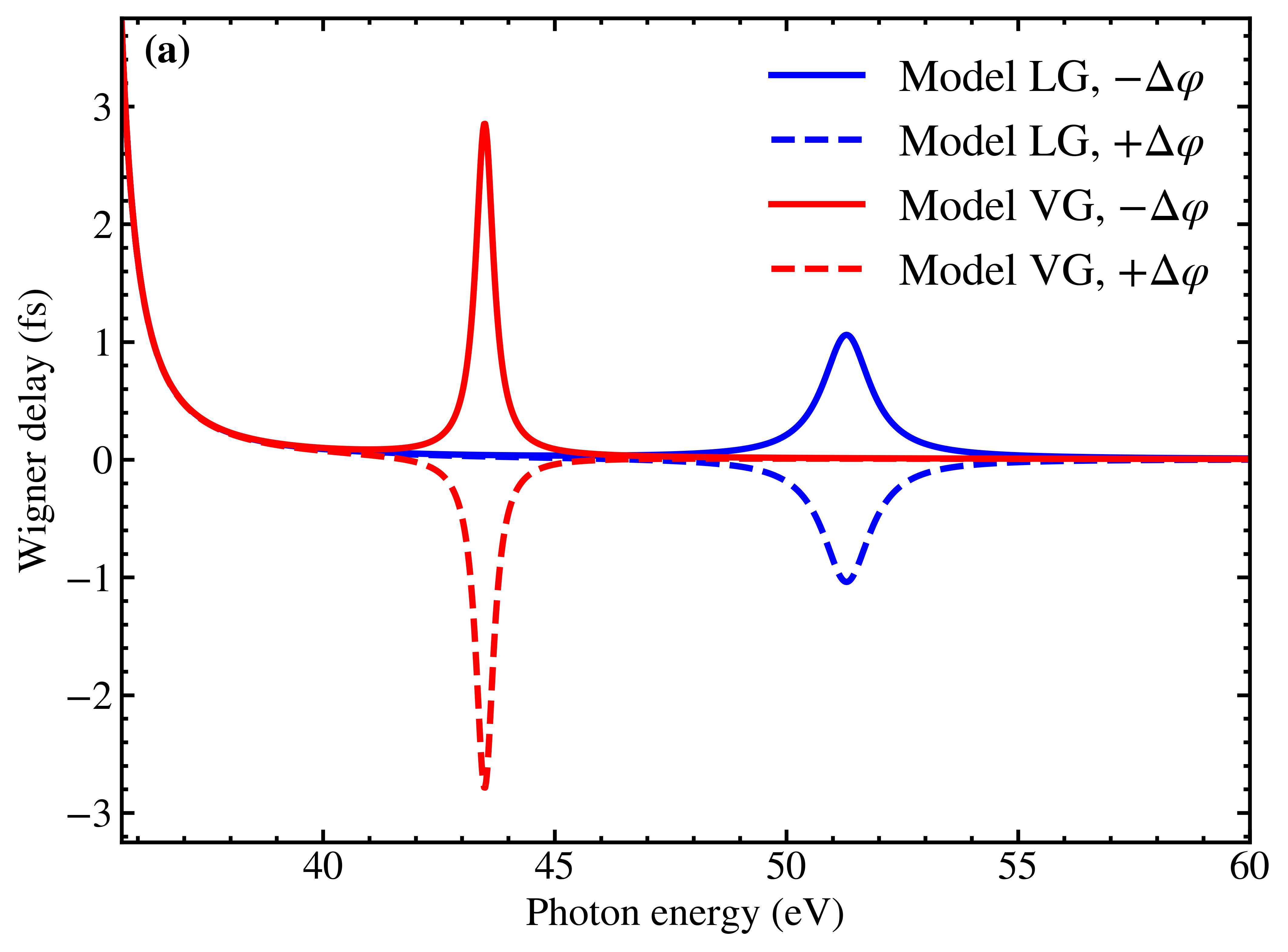} \\ \includegraphics[width=0.49\textwidth]{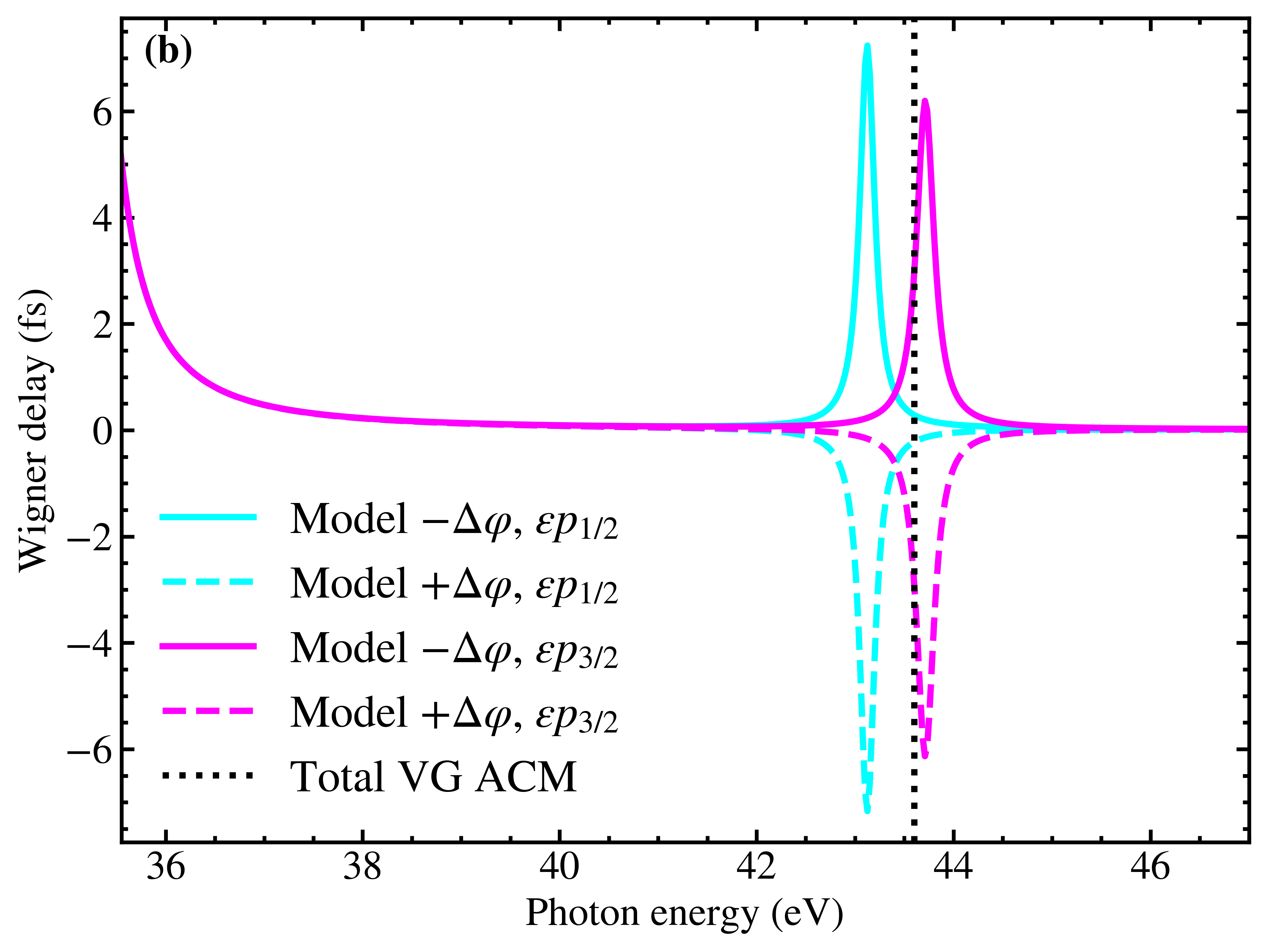} \hfill
    \includegraphics[width=0.49\textwidth]{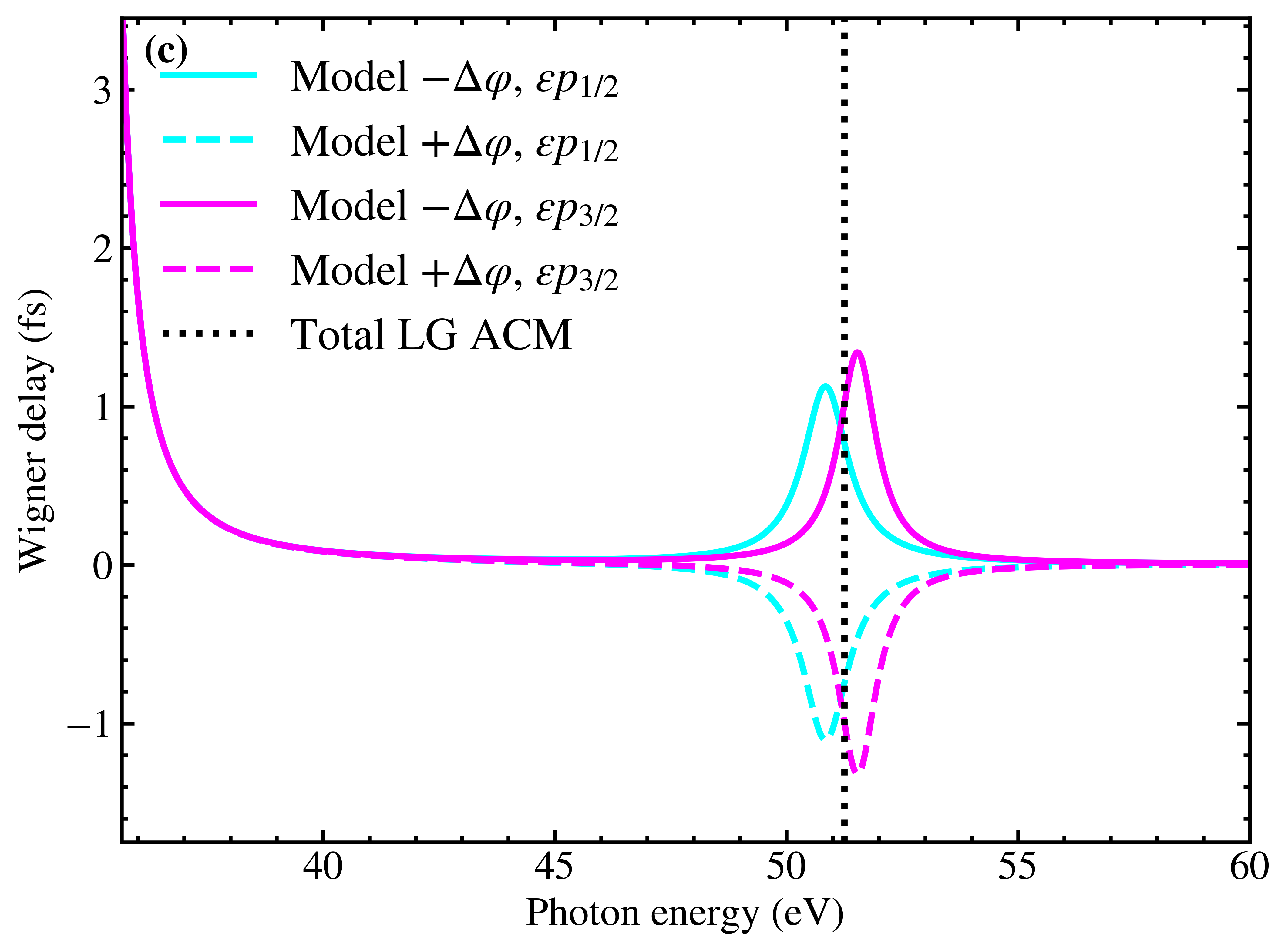}
    \caption{Wigner time delays extracted from the analytical model using parameters obtained by fitting the CS of the model to that of RTDCIS simulations for total $3s$ orbital (a) and relativistic channels in VG (b) and LG (c). The vertical dotted lines in panels (b) and (c) mark the photon energies where the total Wigner delay reaches its extrema for the corresponding gauges shown in panel (a).}
    \label{fig:wigner_lg_vg}
\end{figure}

\subsection{Wigner delays from the relativistic model}
\label{sec:wignerdelay}
Figure \ref{fig:wigner_lg_vg} demonstrates the Wigner time delays as a function of photon energy close to the $3s$ ACM computed by the analytical model for the total $3s$ orbital (a) and separated relativistic channels, (b) and (c). Since the photoelectron Wigner delay is defined as the energy derivative of the phase of the dipole matrix element in Eq. (\ref{eq: dipole model}), it reaches the extrema when $z_{\pm}(\omega)$ passes close to zero, which occurs at the ACM. Therefore, the Wigner delay is almost zero in energies other than ACM, and it reaches high values in both gauges close to their ACM. The only exception is when the photon energy approaches the ionization threshold, and high values of the Wigner delay are observed due to the Coulomb phase. In Fig. \ref{fig:wigner_lg_vg}(a), where the total $3s$ time delay is presented, the curves of VG are sharper and reach high values of $\pm 2.84$ fs, while in LG, the maximum value decreases to $\pm 1.05$ fs with a broader width. It should be noted that the magnitude of the Wigner delay is larger than that recently reported in Ref \cite{sizuo_prl_2025}, where shake-up processes were included. In each gauge, a small relative phase shift from $\Delta \varphi=0$ leads to dramatic changes in the delay around the ACM, flipping the sign of the Wigner delay peak. A positive Wigner time delay implies that the correlated wave packet is delayed, while a negative value implies that it is advanced, compared to an uncorrelated wave packet. The relative phase shift required to switch the sign of the Wigner delay is just $0.08$ rad ($4.6^\circ$) in LG and $0.10$ rad ($5.7^\circ$) in VG, demonstrating the sensitivity of the photoionization time delay to the coherent addition of complex amplitudes at the destructive interference of the ACM. \\
Figs. \ref{fig:wigner_lg_vg} (b) and (c) demonstrate channel-resolved Wigner delays in VG and LG corresponding to the CS shown in Figures \ref{fig:cs_lg_vg}(b) and (c), respectively. In Fig. \ref{fig:wigner_lg_vg} (b), VG exhibits channel-dependence of the Wigner delay peaks with $j=3/2$ appearing at higher energy and lower peak, compared to the $j=1/2$ channel. In LG, Fig. \ref{fig:wigner_lg_vg} (c), the channels also show shifted maxima of the Wigner delay with $\epsilon p_{3/2}$ found at higher energy than $\epsilon p_{1/2}$.  However, the channel with $j=3/2$ has larger peak delays compared to the $j=1/2$ channel, as opposed to VG. Additionally, the separation between the energies of the two relativistic channels in LG is smaller than in the VG case.\\ 
p
\begin{figure}[htbp]
    \centering
    \begin{subfigure}[b]{0.49\textwidth}
        \centering
        \includegraphics[width=\textwidth]{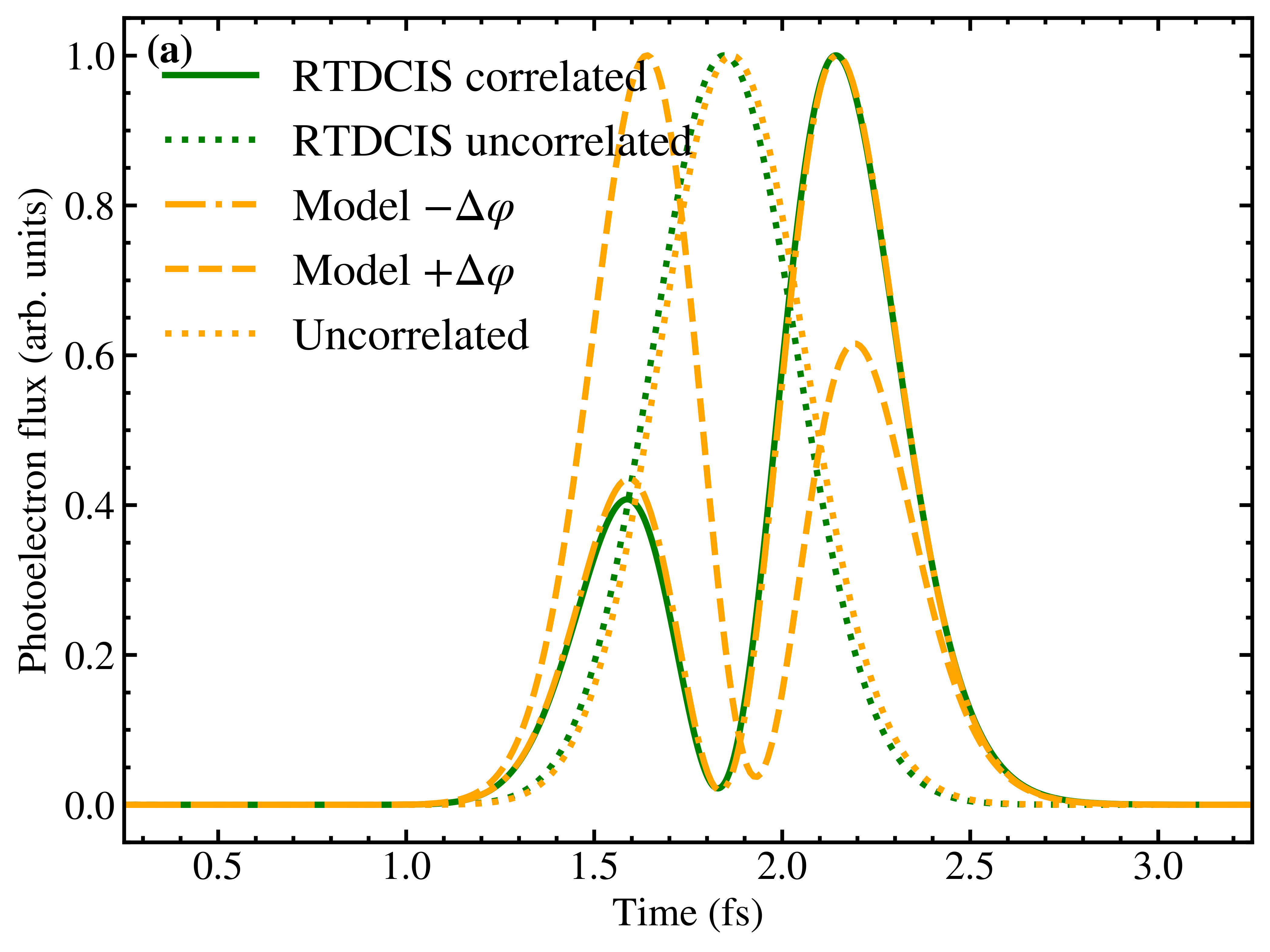}
        
    \end{subfigure}
    \hfill
    \begin{subfigure}[b]{0.49\textwidth}
        \centering
        \includegraphics[width=\textwidth]{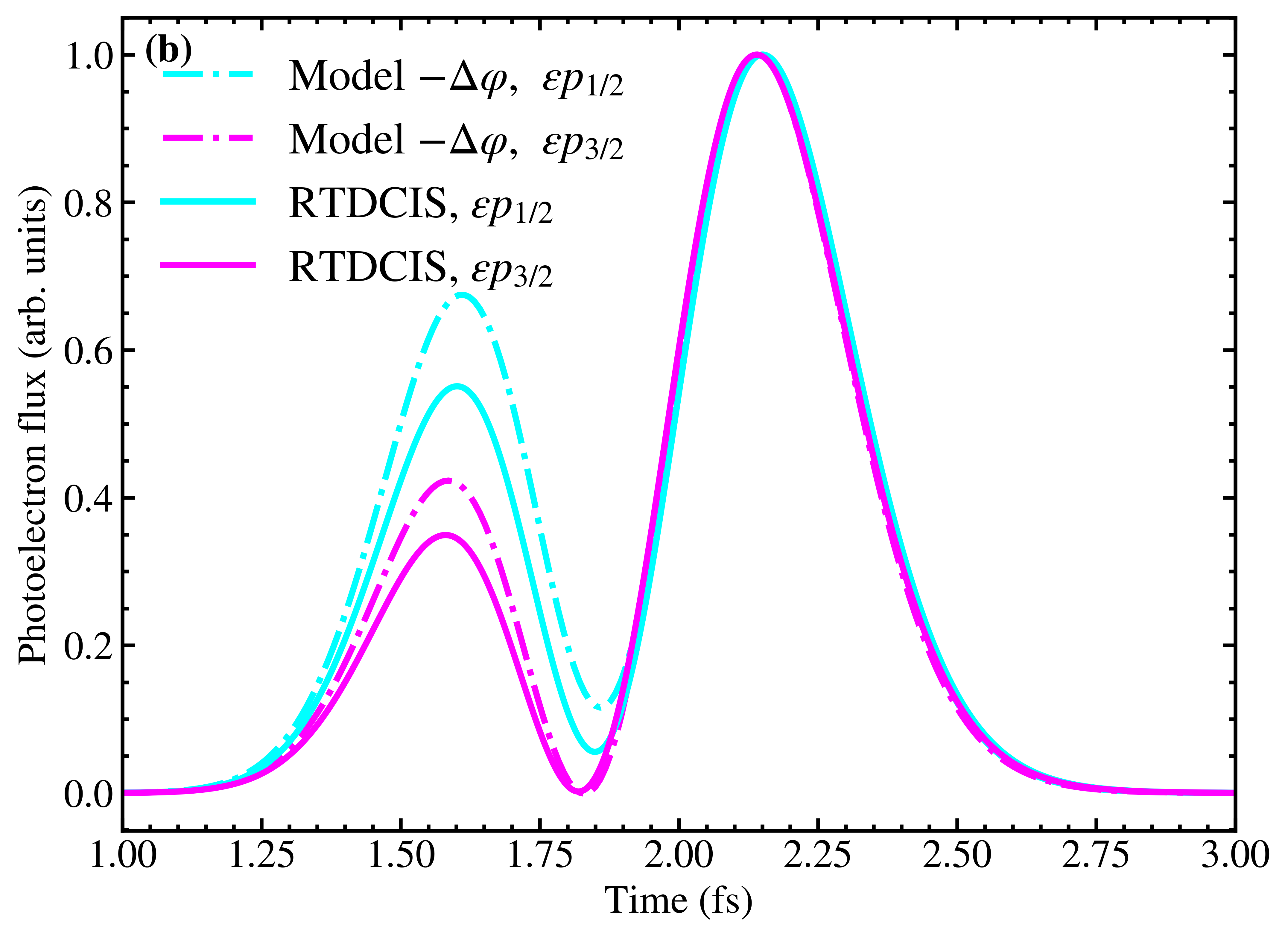}
        
    \end{subfigure}
    
    \caption{(a) Photoelectron fluxes observed at $R \simeq 87.55$ Bohr for LG with $\tau=362.7$ as, ACM photon energy, and $I=10^{12}$ W/cm$^2$ in RTDCIS and the model. (b) channel-resolved ($\epsilon p_{j}$) fluxes corresponding to the RTDCIS correlated case and the model with a negative sign of $\Delta \varphi$ in (a)}
    \label{fig:flux_lg_15}
\end{figure}

\subsection{Time-dependent relativistic flux: Attosecond pulse excitation}
\label{sec:tdflux-as}
Next, we study time-dependent fluxes when the system interacts with an attosecond pulse of $\tau=362.7$ as duration, the intensity of $I=10^{12}$ W/cm$^2$ tuned to the $3s$ ACM of each gauge (i.e. $\omega_0=51.25$ eV for LG and $\omega_0=43.6$ eV for VG). The observation point of photoelectrons is $R\simeq 87.55$ Bohr. Fig. \ref{fig:flux_lg_15} (a) shows the normalized fluxes computed numerically by RTDCIS using Eq. (\ref{eq: rtdcis flux}) (green curves) and the fitted model with Eq. (\ref{eq:flux}) (orange curves). In the numerical results, there are two cases: uncorrelated, where only the $3s$ channel is active in the simulations, and the correlated case with $3s$ and $3p$ activated. Analytical fluxes have three possibilities: uncorrelated and correlated with positive and negative phase shifts. The uncorrelated curves from RTDCIS and the model have a single peak at $1.84$ fs and $1.86$ fs, respectively. However, the correlated cases demonstrate a double-peaked feature; photoelectrons are divided between fast- and slow-moving, with the minimum between them being the manifestation of ACM in the time domain. Slow electrons arrive at the observation surface after ACM, and fast ones arrive before that. The peak of the uncorrelated curve is a suitable time reference, and we can reliably discuss the arrival time of different parts of the correlated flux in comparison to it. 
In the correlated case, when the peak of slow electrons dominates, it means that the total flux arrives later than the reference, and the opposite applies to a dominant fast electron peak. In Fig. \ref{fig:flux_lg_15} (a), the solid and dash-dotted lines agree well; both exhibit a pronounced peak of slow electrons, while the dashed line has more flux at faster electrons. This reveals that to obtain a consistent flux distribution from the model, as compared with LG RTDCIS, $\Delta\varphi$ needs to have a negative sign, and correlated photoelectrons are delayed relative to the uncorrelated case. Such delay is consistent with RPAE calculations of the Wigner delay in the sense that they also exhibit a positive peak value \cite{kheifets_time_2013, Dahlstrom2012PRA}.\\
Figure \ref{fig:flux_lg_15}(b) presents the relativistic channel-resolved fluxes produced by the correlated RTDCIS, along with those computed using the model with a negative correlation phase. The results exhibit only subtle variations depending on the $j$ value of the free electron, $\epsilon p_j$. Each channel reaches its local minimum at slightly different positions, and the slow electrons are dominant. The fast electron peak shows some weak $j$-dependence in the model and RTDCIS. In general, the simple model works quite well for reproducing the \textit{ab initio} results, but it leads to slightly exaggerated peaks for the fast electrons. 
\begin{figure}[htbp]
    \centering
    \includegraphics[width=\textwidth]{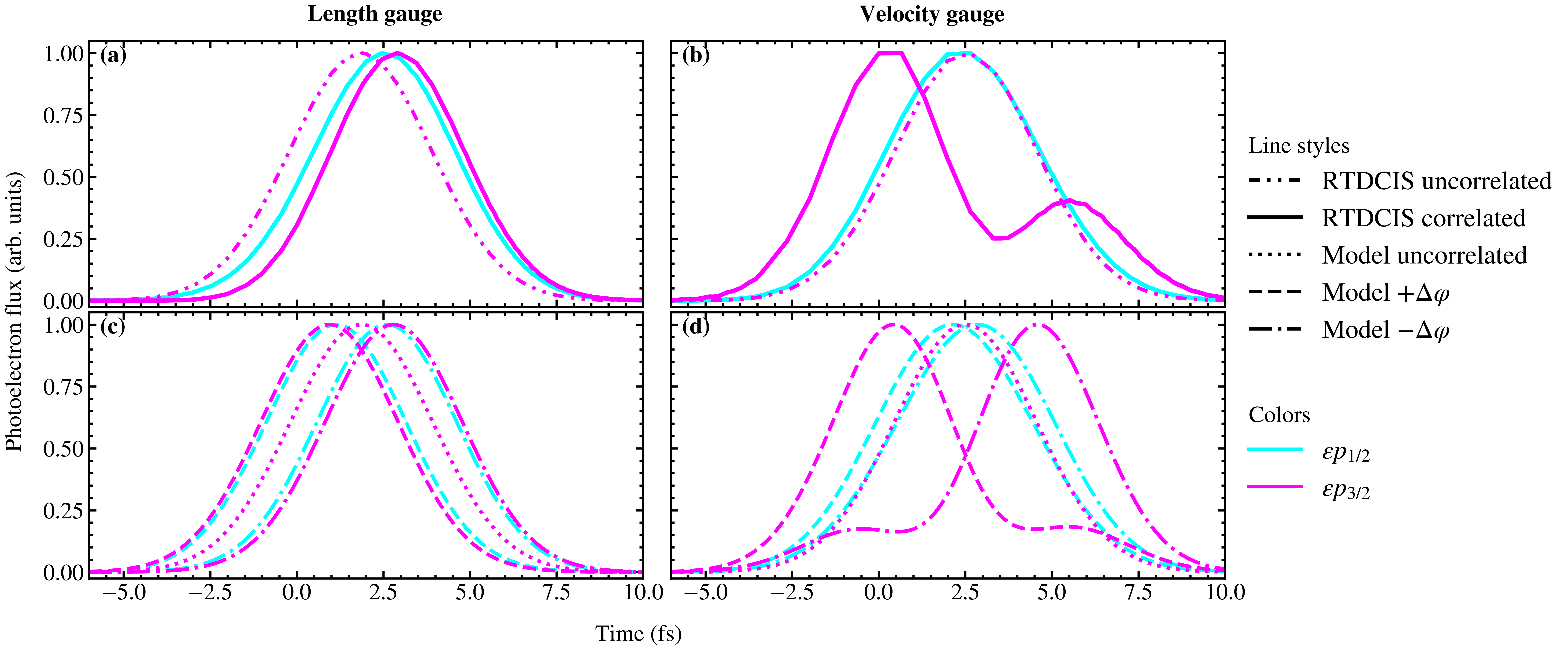}
    \caption{
Channel resolved photoelectron fluxes at $R\simeq 87.55$ Bohr for a pulse of $4.84$ fs duration, intensity of $I=10^{12}$ W/cm$^2$, and the photon energy tuned to the ACM frequency of each gauge. Panels (a) and (b) show the RTDCIS results for LG and VG, respectively, and panels (c) and (d) display the corresponding model calculations. In all panels, curve colors indicate the particle’s angular momentum $\epsilon p_{j}$. The dash–dot–dotted curves show the uncorrelated RTDCIS results, whereas the solid curves show the correlated RTDCIS results. Dotted curves represent the uncorrelated model calculations, while the dashed and dash–dotted curves correspond to the correlated model results with positive and negative phases, respectively.
}
\label{fig:channel_resolved_fluxes_200}
\end{figure}

\subsection{Time-dependent relativistic flux: Femtosecond pulse excitation}
\label{sec:tdflux-fs}
Fig. \ref{fig:channel_resolved_fluxes_200} demonstrates the photoelectron fluxes obtained from the RTDCIS simulations and from the model calculations in both gauges with a pulse duration of $4.84$ fs. In panel (a) for LG, fluxes exhibit a single peak structure, which implies that for this gauge, the employed pulse duration is narrow in energy. In VG, plotted in panel (b), a channel-dependent behavior emerges; the $\epsilon p_{1/2}$ shows a single-peak structure, whereas the $\epsilon p_{3/2}$ has two peaks. In this case, both channels show comparable broadening in Figs. \ref{fig:cs_lg_vg} (b) and \ref{fig:wigner_lg_vg} (b), which indicates that the pulse duration alone can not account for the significant difference between them. This behavior can instead be understood by noting that the total $3s$ ACM lies closer to the $j=3/2$ channel, making its dynamics more strongly influenced, while the $j=1/2$ is more detuned and therefore less affected. Moreover, in LG, the peak of the correlated fluxes occurs later than that of the uncorrelated fluxes, whereas in VG, it appears earlier. This implies that photoelectrons are delayed in LG and advanced in VG.  \\
Figs. \ref{fig:channel_resolved_fluxes_200} (c) and (d) are the corresponding model simulations for LG and VG, each having two possibilities for the sign of correlation phase shift, $\Delta \varphi$. By comparing the upper and lower panels on each side, it is confirmed that the curve with a negative phase shift of the model corresponds well to the correlated RTDCIS result in LG. In contrast, the agreement is found for the positive correlation phase in VG. Physically, the photoelectron is delayed in LG, while it is advanced in VG, which shows that RTDCIS predicts different temporal behavior in the two gauges. The gauge freedom leads to qualitatively different photoelectron dynamics in the time domain --- despite having similarly looking photoionization cross sections. \\
%


\section{Discussion}
\label{sec:disc}
In this section, we discuss the general interpretation of the numerical and analytical results. In Fig. \ref{fig:cs_lg_vg} for VG (b) and LG (c), we saw that the ACM of the relativistic channel with $j=3/2$ appears at a higher photon energy with a smaller depth compared to the $j=1/2$ channel. The corresponding Wigner time delays, depicted in Fig. \ref{fig:wigner_lg_vg} (b) and (c), showed that the VG result has more clearly separated delay peaks compared to the LG result. How the separation of the delay peaks affects the structure of the fluxes depends on the exact pulse duration and the central photon energy. Throughout the presented results, the photon energy is tuned to the ACM of the total CS from $3s$ in each gauge. For VG, the response is more strongly resonant with $j = 3/2$ than with $j = 1/2$, as indicated by the vertical line in Fig.~\ref{fig:wigner_lg_vg}(b). In contrast, for LG the field is more evenly resonant with both $j$-values, as shown by the vertical line in Fig.~\ref{fig:wigner_lg_vg}(c). 

The strong effect of pulse duration is seen by comparing the LG results in Fig. \ref{fig:flux_lg_15} (short pulse of $362.7$ as), with Figs. \ref{fig:channel_resolved_fluxes_200} (a) and (c) (longer pulse of $4.84$ fs). It was observed that the double peak changes to a single peak for the longer pulse. The short pulse is broader in energy and excites both fast- and slow-moving electrons, which separate to different arrival times at the observation point. Moreover, the effect of pulse duration is not the same in the two gauges, as was demonstrated in Fig. \ref{fig:channel_resolved_fluxes_200}, where the LG results (a) showed a single wave packet for both relativistic channels, while the VG results (b) showed different wave packet dynamics, with a persistent splitting of arrival times in the $j=3/2$ channel.  
Finally, based on the analytical model, shown in Fig.~\ref{fig:channel_resolved_fluxes_200}, we could infer from the flux of RTDCIS, that the VG has negative Wigner delay peaks, while LG has positive peaks, at the relativistic ACMs.

\subsection{Role of excitation examined with the Wigner distribution}
\label{sec:caus-Wig}
\begin{figure} 
    \centering
    \includegraphics[width=\textwidth]{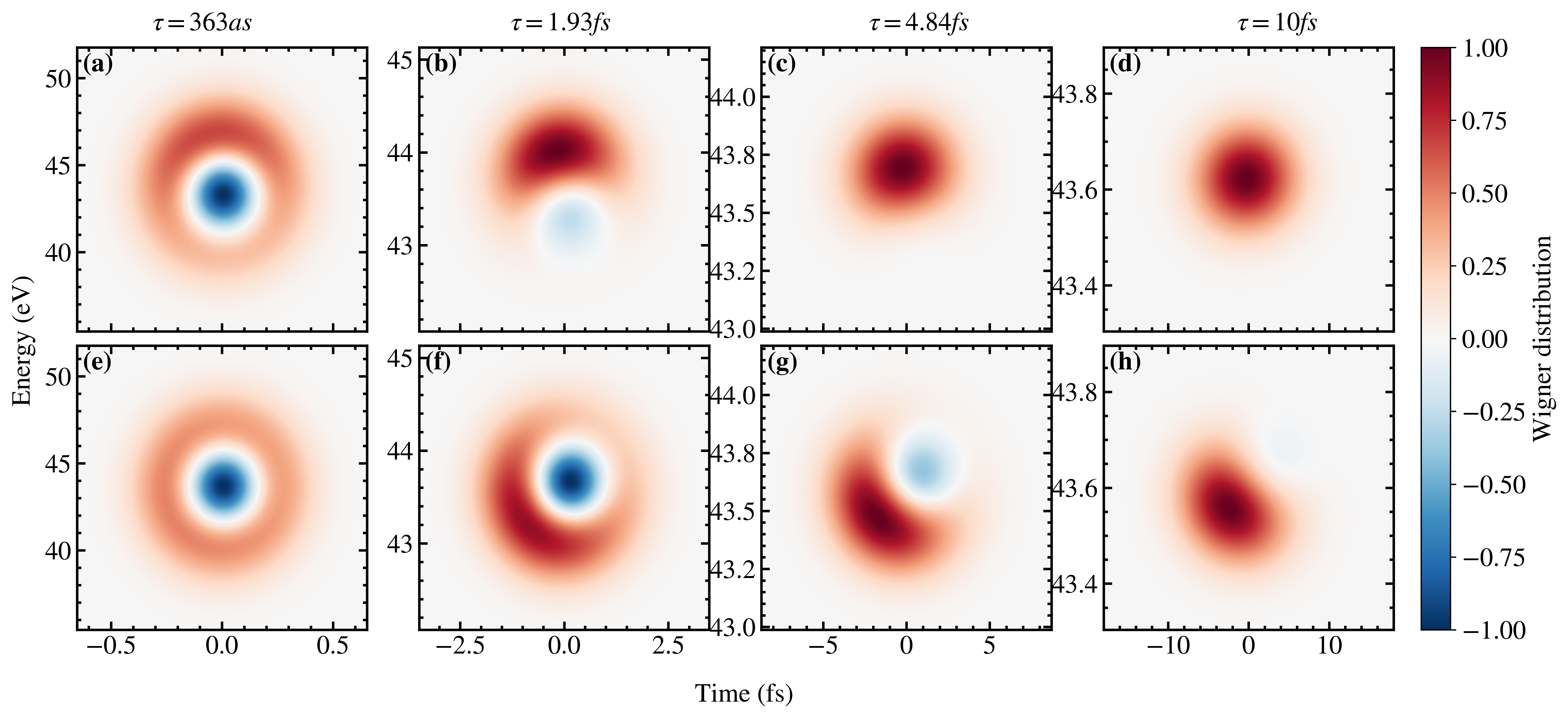}
    \caption{Wigner distributions in VG with positive correlation phase for different pulse durations and relativistic channels: top row corresponds to $j=1/2$ and bottom row to $j=3/2$, pulse durations are $363$ as for panels (a) and (e), $1.93$ fs for panels (b) and (f), $4.84$ fs for panels (c) and (g), and $10$ fs for panels (d) and (h).}
    \label{fig:wigner_dist_vg}
\end{figure}

To further analyze the simultaneous effect of pulse duration on the temporal and energetic structure of the photoelectron, we examine the {\it Wigner distribution} defined in Eq. (\ref{eq:wigner distribution}). Figure~\ref{fig:wigner_dist_vg} presents the Wigner distributions in VG (negative Wigner delays = positive correlation phases) across four pulse durations: $363$~as for panels~(a) and~(e), $1.93$~fs for panels~(b) and~(f), $4.84$~fs for panels~(c) and~(g), and $10.00$~fs for panels~(d) and~(h), for $j=1/2$ (top row) and $3/2$ (bottom row). For the shortest pulse, both relativistic channels display interference, characterized by a central region of negative values surrounded by positive values. The Wigner distribution is reminiscent of a {\it Schrödinger kitten} state in quantum optics, {\it c.f.} Ref.~\cite{lewenstein_generation_2021}, because it consists of a coherent superposition of two photoelectron wavepackets: one slow and the other fast. As the pulse duration increases to the intermediate values, the interference becomes clearly channel-dependent: the $j=3/2$ relativistic channel maintains a clear interference structure, similar to a {\it Schrödinger cat} state, whereas the other exhibits a nearly single peak with a more Gaussian distribution. For longer pulses, the interference is gradually reduced with the $j=1/2$ channel showing a positive distribution at a higher energy than the $j=3/2$ channel. This effective energy shift between the relativistic channels can be understood as a result of the opposite slopes of the relativistic CS. Note also that the negative temporal shift of the $j=3/2$ channel is consistent the negative Wigner delay of the ACM in VG. 

Our analysis demonstrates that the temporal features depend sensitively on the exact pulse duration and resonance energy for each $j$ quantum number. 
For this reason, we now define a {\it weighted delay}:
\begin{equation}
    \bar\tau=\frac{\int dt\int d\omega \, t\, {\cal W}(\omega,t)}{\int dt\int d\omega \, {\cal W}(\omega,t)}
    =
    \frac{\int dt \, t \, |\tilde{\cal Z}(t)|^2}
    {\int dt \, |\tilde{\cal  Z}(t)|^2},
    \label{eq: tau_bar_wigner}
\end{equation}
and a {\it weighted energy}:  
\begin{equation}
\bar\omega=\frac{\int dt\int d\omega \, \omega \, {\cal W}(\omega,t)}{\int dt\int d\omega \, {\cal W}(\omega,t)} = 
    \frac{\int d\omega \, \omega \, |{\cal Z}(\omega)|^2}
    {\int d\omega \, |{\cal Z}(\omega)|^2},
\label{eq: omega_bar_wigner}
\end{equation}
using the effective dipole ${\cal Z}(\omega)$ and its Fourier transform $\tilde{\cal Z}(t)$. In contrast to conventional Wigner delays, weighted delays take into account the temporal duration of the excitation thanks to the bandwidth filter in the effective dipole.    
Numerical values of the weighted quantities are given in Table~\ref{tab:weight}. In both VG and LG, as $\tau$ becomes longer, $\bar{\omega}$ increases (decreases) for $j=1/2$ ($j=3/2$) towards $\omega_{0}$, so for each pulse duration, the value of $\bar{\omega}$ is larger in $j=1/2$ than in $j=3/2$. The case of $\tau=0.363$ fs and $j=1/2$ in LG is the one exception to this simple rule, due to the large bandwidth covering the entire ACM.

\subsection{Nature of causality at the ACM}
\label{sec:dir-cau}
Examining $\bar{\tau}$ in Table~\ref{tab:weight}, the values in VG are mostly negative, while they are positive in LG, which is in qualitative agreement with the Wigner delays for the two gauges, $\tau_\mathrm{EWS}$ in Table \ref{tab:delay_contributions}. 
There is, however, one anomaly for attosecond pulses, see $\bar \tau$ for $\tau=363$\,as in Table \ref{tab:weight}, where the weighted delay in VG becomes positive, {\it i.e.} the photoelectron becomes delayed rather than advanced. This raises the question of whether the Wigner delay approach is inadequate for attosecond pulses? 
In general, $|\bar{\tau}|$ increases in both gauges and channels as $\tau$ gets longer. The values of $|\bar{\tau}|$ for $j=3/2$ are much larger than $1/2$ in VG for all pulse durations. In LG the two weighted delays of $3/2$ and $1/2$ are more comparable and also increase with the pulse duration. We expect that the weighted delay at a given central frequency will tend to the Wigner delay as the excitation window increases, which is indeed found in LG. In VG, the sharp variation of the cross section may still affect the weighted delay.  
We interpret this behavior as a consequence of {\it direct causality}: The photoelectron wave packet will not be created outside the main excitation window of the pulse: $|\bar\tau|<\tau/2$. All weighted delays in Table \ref{tab:weight} respect this direct causal argument, providing a simple explanation for why the absolute values of the weighted delays decrease as the excitation pulse gets shorter.

To derive this result, we may consider the Fourier transform of the effective dipole as the response for photoionization in the time domain, $\tilde{\cal Z}(t)=\mathrm{FT}[{\cal Z}(\omega)]$. If the dipole is approximated as constant in the energy domain, $z(\omega)\approx a_0$, then the response function would simply be the Fourier transform of the spectral envelope, ${\cal Z}(t)=\mathrm{FT}[a_0g(\omega)]=a_0\tilde g(t)$, which for a Gaussian would be also a Gaussian in time. If, instead, the dipole is a linear polynomial in energy, $z(\omega)\approx a_0+b_0\omega$, then the temporal response will be a linear combination of the temporal envelope and its time derivative, ${\cal Z}(t)=\mathrm{FT}[(a_0+b_0\omega)g(\omega)]=a_0\tilde g(t)+i b_0\dot{\tilde g}(t)$, which for Gaussian envelopes imply even (single peak) and odd (double peak) parts that shape the response. Qualitatively, this simple argument explains the asymmetric double peak structures found in Figs.~\ref{fig:flux_lg_15} and \ref{fig:channel_resolved_fluxes_200}. Higher-order polynomial expansion leads to further reshaping of the temporal response, but the net result will be localized to the initial pulse, which we refer to as direct causality.     

More generally, causality forbids consequence before action: $\tilde z(t)=0$ for $t<0$, which any $z(\omega)$ that is analytical in the upper energy complex plane fulfills. 
If the dipole contains a pole in the complex plane, such as a Breit-Wigner resonance: 
\begin{equation}
z_\mathrm{BW}(\omega)=\frac{1}{(\omega-\Omega+i\Gamma/2)}=\frac{2/\Gamma}{(\epsilon+i)},
\label{eq:bw}
\end{equation}
where $\epsilon(\omega)=(\omega-\Omega)/(\Gamma/2)$, then the causality argument widens to $-\tau/2<\bar\tau<1/\Gamma$, where $\tau<1/\Gamma$. This extension of the emission time is due to the convolution, 
\begin{equation}
\tilde{\cal Z}_{\mathrm{BW}}(t)\sim \int dt' \tilde g(t-t')\exp(-i\Omega t')\exp(-\Gamma t'/2)\theta(t'), 
\label{eq:fano-conv}
\end{equation}
which gives the system memory to respond {\it after} the direct causal window of the incident pulse. Due to causality, the complex conjugate of the resonance dipole, $z_\mathrm{BW}^*(\omega)$ is not an allowed dipole model because it would contain a pole in the upper-half complex plane. Thus, there is no phase ambiguity: the Wigner delay peak at the BW resonance is {\it always} positive --- the resonance temporarily traps the photoelectron. 
In the case of Fano resonances, 
\begin{equation}
z_\mathrm{F}(\omega)=\frac{(q+\epsilon)}{(\epsilon+\mathrm{i})}, 
\label{eq:fano}
\end{equation}
where the Fano parameter is real: $q\in \mathbb{R}$, and a pole is again found in the lower-half complex plane. The response is causal like in the BW case. 
Physically, a resonant path now interferes with a direct path, creating a dipole phase with two distinct contributions: 
\begin{equation}
\arg(z_\mathrm{F})=\arg(q+\epsilon)-\arg(\epsilon+\mathrm{i}), 
\label{eq:fanophase}
\end{equation}
where the former is a discontinuous phase jump by $\pi$ at $\epsilon=-q$ due to interference, while the latter is identical the BW case. There are two allowed signs of the real $q$-parameter, which leave their mark on the shape of the CS and the phase of the dipole  \cite{fano_effects_1961}. 
If the Fano parameter is taken to be a complex number: $q\in \mathbb{C}$, then the discontinuous phase shift is smoothed out, with locally decreasing (increasing) phases for $\mathrm{Im}~q>0$ ($\mathrm{Im}~q<0$), c.f. Refs.~\cite{jimenez-galan_modulation_2014,kotur_spectral_2016}. It has been found that the Fano profile with complex $q$ fits well to the ACM \cite{Hans_Jakob_2024}, and that the additional {\it minimum-phase condition}, {\it i.e.} $\mathrm{Im}~q>0$, then yields a negative Wigner delay locally at the ACM in excellent agreement with experiments \cite{Alexandridi2021PRR,sizuo_prl_2025}. It should be added that the other choice, $\mathrm{Im}~q<0$, yields an identical cross-section, similar to the degeneracy found in our ACM model, and that both signs of the $\mathrm{Im}~q$ respect causality.   

The ACM is not a simple Breit-Wigner resonance, nor is it a complex Fano resonance. It is a smooth interference phenomenon of two coupled continua: a problem with no poles in the complex plane. 
The minimum phase-condition fails when the dipole has a zero in the upper-half complex plane, but such an event will not violate causality. This notion is confirmed by our RTDCIS simulations because: 
\textit{i.} the behavior of $\bar \tau$ in Table~\ref{tab:weight} is consistent with direct causal behavior; and 
\textit{ii.} we observe both signs of the Wigner delay at the ACM in our {\it ab initio} simulations. Thus, based on only the CS, our results imply that it is not possible to know the correct sign of the Wigner delay, or if the winding number is 0 or 1, as formulated by Ji {\it et al.} using the Kramers-Kronig relations with the CS in Ref.~\cite{Hans_Jakob_2024}. A positive Wigner delay at a minimum of the CS could be possible, but it then should be considered an exotic case that breaks the minimum-phase condition. 
While the advance of the Wigner delay from the $3s$ orbital in experiments \cite{Alexandridi2021PRR,sizuo_prl_2025} follows the minimum-phase condition, attosecond physics holds the promise of finding a real system exhibiting more exotic dynamics in the future.

\begin{table}[h]
\centering
\caption{Weighted delays and energies from ${\cal W}(\omega,t)$ for VG and LG with central frequencies $\omega_0^{\mathrm{LG}}=51.25$ eV and $\omega_0^{\mathrm{VG}}=43.6$ eV, respectively.}
\begin{tabular}{c c | cc | cc}
\br
& & \multicolumn{2}{c|}{LG} & \multicolumn{2}{c}{VG} \\
\mr
Channel & $\tau$ (fs) 
& $\bar\omega$ (eV) & $\bar\tau$ (as)
& $\bar\omega$ (eV) & $\bar\tau$ (as) \\
\mr

\multirow{4}{*}{$j=1/2$}
& $0.363$ & $51.26$ & $90.46$ & $43.98$ & $54.71$ \\
& $1.93$ & $51.43$ & $595.11$ & $43.98$ & $-144.21$ \\
& $4.84$ & $51.29$ & $735.99$ & $43.70$ & $-229.93$ \\
& $10.0$ & $51.26$ & $762.40$ & $43.62$ & $-250.14$ \\
\mr

\multirow{4}{*}{$j=3/2$}
& $0.363$ & $50.01$ & $77.33$ & $42.92$ & $67.20$ \\
& $1.93$ & $51.05$ & $672.20$ & $43.40$ & $-359.48$ \\ 
& $4.84$ & $51.21$ & $933.18$ & $43.48$ & $-1418.97$ \\
& $10.0$ & $51.24$ & $989.51$ & $43.56$ & $-2356.65$ \\
\br
\end{tabular}
\label{tab:weight}
\end{table}

\subsection{Interpretation of the delay of RTDCIS flux}
\label{sec:interp}
The {\it average delay} of the RTDCIS is computed as
$$
\bar \tau_\mathrm{RTDCIS} = \frac{\int dt\, t {\cal J}(t) }{ \int dt \,{\cal J}(t)}
$$
where ${\cal J}(t)$ is the flux in Eq.(\ref{eq: rtdcis flux}). 
Numerical values for average delays are presented in Table \ref{tab:delay_contributions}. 
To interpret the origin of the delays observed in the RTDCIS wave packets, we revisit the phase of a photoelectron wave packet, Eq. (\ref{eq:wave packet}) at a given photon energy, $\omega$, following Dahlström, L'Huillier and Maquet in Ref.~\cite{Dahlstrom2012JPB}: 
\begin{equation}
\arg [\Psi_\pm(\omega; r, t)] \sim
\underbrace{k(\omega)r -\frac{k(\omega)^2}{2}t}_{\mathrm{free \ propagation}} 
+\underbrace{\frac{\ln[2k(\omega)r]}{k(\omega)}}_{\mathrm{long-range\ phase}}+\underbrace{\phi[k(\omega)]}_{\mathrm{short-range \ phase}}.
\label{eq: phases}
\end{equation}
The stationary contribution of the first two terms: 
$\frac{d}{dk}(kr-\frac{k^2}{2}t)=0$, implies classical motion of a free electron, $r=k(\omega)t$, where the speed of the electron is $k(\omega)=\sqrt{2(\omega-I_p)}$ \cite{DECARVALHO200283}. The time it takes to propagate to $r=R$ is $\tau_{\mathrm{free}}=R/k_0$ assuming a central frequency of excitation, $\omega_0$. The third phase term diverges logarithmically and originates from the long-range nature of the Coulomb potential \cite{BetheSalpeter1957}, which leads to a time delay $\tau_{\mathrm{long}}\approx 1/k^3_0 [1-\ln(2k_0R)]$ that diverges for large observation points $R=k_0\tau_{\mathrm{free}}$ \cite{Dahlstrom2012JPB}. Finally, the fourth phase term is the short-range phase contribution, which includes the asymptotic Coulomb phase \cite{Dahlstrom2012JPB} and short-range scattering phase corrections due to correlation effects \cite{Dahlstrom2012PRA, kheifets_time_2013}, which are both included in the complex dipole matrix element, see Eq.~(\ref{eq: phases}). Our model is not exact, but it provides us with a tool to interpret the nature of the photoionization dynamics. While it is the energy derivative of the short-range part that is used to compute the Wigner delay, $\tau_{\mathrm{EWS}}$, both free propagation and long-range effects will affect the actual arrival time of the wave packet due to causality: excitation before propagation \cite{Dahlstrom2012JPB}. Table \ref{tab:delay_contributions} shows the values of the different terms contributing to the arrival time of the wave packet, as well as the average delay of the RTDCIS fluxes using the same pulses as in Fig.~\ref{fig:channel_resolved_fluxes_200}. 

\begin{table}[h]
\centering
\caption{Values of different contributions to the arrival time of fluxes from Eq.~(\ref{eq: phases}) together with the values obtained by RTDCIS simulations for velocity (VG) and length (LG) gauges. The observation radius is $R\simeq 87.55$ Bohr, and the central photon energy is tuned to the ACM of each gauge 
with $\tau=4.84$ fs. All values in the table are in units of attoseconds.}
\begin{tabular}{lc|cccc|c|cc}
\br
\textbf{} & Channel &
\textbf{$\tau_{\mathrm{free}}$} & \textbf{$\tau_{\mathrm{long}}$} &
\textbf{$\tau_{\mathrm{EWS}}$} & \textbf{$\tau_{\mathrm{total}}$} &
\textbf{$\bar\tau_\mathrm{RTDCIS}$} & $\Delta_\mathrm{EWS}$ & $\bar\Delta$ \\
\br

\multirow{3}{*}{LG}
    & Uncorr.   & $1938.42$ & $-78.90$ & $10.57$ & $1870.09$ & $1845.77$ & $-24.32$ &  \\
    & $j=1/2$   & $1936.03$ & $-78.64$ & $770.53$ & $2627.92$ & $2496.53$ & $-131.39$ & $-96.85$ \\
    & $j=3/2$   & $1940.81$ & $-79.17$ & $1007.36$ & $2869.00$ & $2964.80$ & $95.80$ & $169.98$ \\
\br

\multirow{3}{*}{VG}
    & Uncorr.   & $2665.29$ & $-189.76$ & $32.32$ & $2507.85$ & $2523.77$ & $15.92$ & \\
    & $j=1/2$   & $2649.91$ & $-186.77$ & $-222.30$ & $2240.84$ & $2488.05$ & $247.21$ & $254.84$ \\
    & $j=3/2$   & $2684.11$ & $-193.46$ & $-2883.86$ & $-393.21$ & $1621.08$ & $2014.29$ & $549.4$ \\
\br
\label{tab:delay_contributions}
\end{tabular}
\end{table}

In Table \ref{tab:delay_contributions}, the $\tau_{\mathrm{free}}$ and $\tau_{\mathrm{long}}$ are computed using the central frequency, $\omega_0$, for the uncorrelated case and the weighted frequencies, $\bar{\omega}$, for the relativistic channels. The values of $\tau_{\mathrm{free}}$ and $\tau_{\mathrm{long}}$ are larger in VG, as expected, since it has a lower $\omega_0$, so the photoelectrons are slower and take more time to arrive at the observation point. In both gauges, it takes a longer time for the $3/2$ channel to arrive compared to the $1/2$ due to the opposite slopes of the ACM at the central frequency. Examining $|\tau_{\mathrm{EWS}}|$, which is computed at the central frequency of the pulse, uncorrelated values are smaller than the correlated ones in all cases. In VG, $j=3/2$ has a significantly larger value than $j=1/2$; which is expected because $\omega_0$ is more tuned to the $j=3/2$ ACM. Also, in VG, correlated cases have negative values, indicating that the correlated wave packet has advanced compared to the uncorrelated one. The column marked as $\tau_{\mathrm{total}}$ is the summation of the first three columns and corresponds to the expected arrival time of the fluxes computed by the model. All cases show a delay (a positive total arrival time), except for $j=3/2$ in VG, where an advancement of around $393$ as is predicted. Comparing the corresponding flux in Fig. \ref{fig:channel_resolved_fluxes_200} (d), we clearly see that the wave packet should be delayed, as expected from causality. What is causing this discrepancy between the Wigner delay interpretation and the RTDCIS flux? 
All values in the column of $\bar\tau_\mathrm{RTDCIS}$ are positive. 
The column $\Delta_{\mathrm{EWS}}$ shows the difference between $\bar\tau_\mathrm{RTDCIS}$ and $\tau_{\mathrm{total}}$; as expected, the difference is relatively small compared to the total arrival time of the flux, except for the VG $j=3/2$ case. In this case, the discrepancy exceeds the average arrival time ($124\%$). This shows that the Wigner time delay concept does {\it not} hold for these modest pulse parameters in the $j=3/2$ channel. Further, we note that the Wigner delay violates the causal relation, because $\tau_\mathrm{EWS}<-\tau/2$. Such obvious violation of causality is not found in any of the other cases in Table \ref{tab:delay_contributions}.  
To improve the predictions of the model, by taking into account causality, we use the weighted delay, $\bar{\tau}$ given in Table \ref{tab:weight}, instead of ${\tau}_\mathrm{EWS}$. The last column, $\bar{\Delta}$ gives the difference between RTDCIS arrival time and the updated model with the new change. We find that using the weighted delay improves the agreement with RTDCIS in the VG $j=3/2$ channel, where the difference is now smaller than the arrival time of RTDCIS. In the other cases, it is not found to affect the result much on the scale of the arrival time of RTDCIS. The fact that the weighted delay does not generally improve the comparison with RTDCIS is probably due to other reasons, such as the limitations of our analytical model for the relativistic case (recall the wave packets in Fig.~\ref{fig:channel_resolved_fluxes_200}). 
For the uncorrelated case, the asymptotic wave packet picture works well with only a small difference ($\approx -24$ as) between the RTDCIS and the model; the size of this error is on the same order as the short-range phase corrections in the uncorrelated case ($\approx -35$ as) \cite{kennedy_photoionization_1972}, which are not considered in our analytical model. 




\section{Conclusion}
\label{sec:concl}

In this work, we investigated ultrafast wave packets and their time delays in photoionization from the $3s$ orbital in argon using RTDCIS in both LG and VG. More precisely, we showed that the Wigner delay is positive for LG and negative for VG. 
Further, we studied the effect of spin-orbit coupling on the ACM for the different relativistic channels: $j= 1/2 $ and $3/2$ of the photoelectron. We showed that the $j$-quantum number leads to different temporal structure of the emitted wave packets and used Wigner distributions to interpret the results. 
We found that Schrödinger cat states are created at the ACM, which appear in the temporal domain as double peak structures at the observation point (a superposition of slow and fast electrons).  

More generally, our work demonstrates that different signs of Wigner delays can emerge from the same physical system, modeled at the same level of many-body theory, but with different choices of gauge to describe its interaction with the external fields. We are not proposing that VG is a better gauge for this type of simulation, despite it having the correct sign of the Wigner delay compared to experiments \cite{Alexandridi2021PRR,sizuo_prl_2025}, because it is known that more exact descriptions of the electron-electron correlation effects are required to describe the CS properly. Therefore, the most important point of our work is that it deepens our understanding of causality in ultrafast photoionization, showing that both positive and negative Wigner delays are theoretically possible at the ACM, because neither leads to contradictions with causality in numerical simulations, or when properly weighted delays, obtained from Wigner distributions, are used to interpret the dynamics.     


\ack
We thank Eva Lindroth, Philipp Demekhin, Felipe Zapata, Asimina Papoulia, and Jimmy Vinbladh for discussions. JMD acknowledges support from the Knut and Alice Wallenberg Foundation: 2024.0212 and the Swedish Research Council: 2024-04247.
\section*{References}
\printbibliography[heading=none]

\end{document}